\documentstyle[floats,prl,aps,psfig]{revtex}		

\psfigurepath{.:/grest1/mdlacas/Blend}

\newlength{\refspace}
\begin{document}
\draft



\wideabs{
\title{Capillary-Wave and Chain-Length Effects at Polymer/Polymer Interfaces}

\author{Martin-D. Lacasse, Gary S. Grest, and Alex J. Levine}

\address{
Corporate Research Science Laboratories\\
Exxon Research and Engineering Co., Annandale, New Jersey 08801}
\date{\today}

\maketitle{%
\begin{abstract}
A continuum-space bead-spring model
is used to study the phase behavior
of binary blends of homopolymers and
the structure of the interface between
the two immiscible phases.
The structure of the interface
is investigated as a function of immiscibility,
chain length, and system size.
Capillary waves are observed and their measurement allows us
to determine the surface tension $\gamma$.
We propose a more universal method of measuring the
interfacial width in terms of second moments of
the different contributions to the first derivative
of the interfacial profile. Predictions of this method are
directly verified.
The effect of chain length on the surface tension
is also studied.
\end{abstract}
}{
\pacs{PACS numbers: 61.41+e, 64.60.Cn, 61.25.Hq}
}
}
\narrowtext


The mixing of two different homopolymer species, say $A$ and $B$,
in the molten state often results in a system of two immiscible phases.
The structure of the interface between the phases
has been the object of several theoretical studies\cite{meanfield,helfand}
which use as a starting point the Flory-Huggins free energy
of the blend and its well-known associated interaction
parameter $\chi_{AB}$\cite{Flory_53}.
These theories predict that the intrinsic order parameter
profile of an interface located at $z_o$
is well-approximated by\cite{definition}
\begin{equation}
\label{eq:profile}
\psi(z) = \psi_o \tanh[2(z-z_o)/w_o],
\end{equation}
where $\psi(z) \equiv (\rho_A(z) - \rho_B(z))/(\rho_A(z) + \rho_B(z))$,
$\rho_I$ is the number density
of species $I$, and $\psi_o$ is the bulk value of the
order parameter. According to the same theories\cite{helfand},
the intrinsic interfacial width $w_o$
is predicted to follow $w_o = a (6\chi_{AB})^{-1/2}$,
while the surface tension $\gamma = (k_BT/a^2)(\chi_{AB}/6)^{1/2}$.
Here, $a$ is the statistical segment length,
$k_B$ is the Boltzmann constant, and $T$ is the temperature.

On the experimental side, neutron reflectivity
experiments\cite{experiments} have measured
a value of $w_o$ that is systematically and substantially
larger than that predicted theoretically
from the $\chi_{AB}$ associated to these blends.
More recent experiments\cite{sferrazza97} seem to support
the proposed explanation\cite{experiments} that the presence
of capillary waves\cite{capillary} at the interface
could be responsible for this discrepancy.

Of similar importance is the effect of chain length $N$
on the interfacial width, since theoretical treatments are
mostly in the limit of infinite $N$. Some finite-chain-length
corrections have been proposed~\cite{chaineffects} but none
of these have been tested systematically.

With these motivations in mind,
a simple continuous-space (CS) model has been developed\cite{model}
in order to study binary blends of homopolymers.
It is the purpose of this Letter
to present results, as derived from this model, for
capillary-wave and chain-length effects at the interface
of immiscible homopolymer blends.
Our approach differs significantly from previous numerical studies
which use lattice models almost exclusively\cite{Binder_95}.
For systems with interfaces, CS models have interesting features:
besides their inherent spatial isotropy and the absence of pinning,
they offer a simple way to determine the surface tension from
the measured pressure tensor\cite{Hill_86}.
Moreover, provided the forces are short-ranged, theoretical
work\cite{robert83} suggests that interfaces
in the continuum exhibit no roughening transition.

The CS model used here has been described in more detail
elsewhere\cite{model}.
In this model, the polymer chains are represented by attaching $N$
soft spherical beads of mass
$m$ using a finitely extensible spring potential.
The softness of the beads, or mers, is set by the interaction potential.
A binary system is built
by constructing a large number $M$ of such chains,
of given type $A$ or $B$, and
enclosing them in a virtual box having the desired boundary conditions.

We use molecular dynamics (MD) as the simulation algorithm, supplemented
by occasional Monte Carlo (MC) type exchanges in order to improve
the sampling of phase space.
The time evolution of the coordinates of all chains is resolved through
Newton's equation of motion (EOM), integrated
using a velocity-Verlet algorithm\cite{Allen_87}.
The motion of the chains is coupled to a heat bath acting through a weak
stochastic force ${\bf W}(t)$ and a corresponding viscous damping term
with friction coefficient $\Gamma$. Besides improving the diffusion
of the system in phase space, this coupling has the practical advantage
of stabilizing the numerical calculation, especially in the
presence of the energy fluctuations induced by the MC exchanges.
Including these terms, the EOM of mer $i$ reads
\begin{equation}
m\frac{d^2\bbox{r}_i}{dt^2} = -\bbox{\nabla}_i U - 
m\Gamma\frac{d\bbox{r}_i}{dt} + {\bf W}_i(t).
\end{equation}
The last term ${\bf W}_i(t)$ is a white noise having an average strength
determined by temperature and the friction coefficient through
the fluctuation-dissipation theorem. In the viscous drag term,
$\Gamma$ has to be carefully chosen
to avoid overdamping
so that the motion of the mers be dominated by their inertia.
Finally, the conservative force term derives from a potential energy
$U$ having two contributions:
the interaction potential $U_{IJ}$
acting between all mers, responsible for excluded volume effects,
and an attractive potential holding adjacent
mers along the chain $U^{ch}$.
For the first contribution in
a mixture of two homopolymer species,
the interaction potential $U_{IJ}(r_{ij})$
between beads $i$ and $j$
of types $I, J = \{A, B\}$ separated by a distance
$r_{ij}$ is taken as the repulsive core of a central-force
Lennard-Jones (LJ) 6:12 potential,
\begin{equation}
U_{IJ}(r_{ij})
= 
4\epsilon_{IJ}
\left[
\left(\frac{\sigma}{r_{ij}}\right)^{12}
        -\left(\frac{\sigma}{r_{ij}}\right)^6
        + \frac{1}{4}
\right]
,
\label{eq:LJ}
\end{equation}
for $r_{ij} < r_c = 2^{1/6}\sigma$ and zero otherwise.
Instead of using the Lorentz-Berthelot mixing rule, we choose
$\sigma = \sigma_A = \sigma_B$, and
$\epsilon_{IJ} = (1+\delta_{IJ}\delta)\epsilon$, where
$\delta \geq 0$ is a small parameter controlling the miscibility
and $\delta_{IJ}$ is the Kronecker delta.
Our choice in energy parameters
can be seen as a special (symmetric) case of
$
\delta=[\epsilon_{AB}
-\frac{1}{2}(\epsilon_{AA}+\epsilon_{BB})]/\epsilon,
$
which represents the kind of interactions
used in simple lattice systems such as in the Flory-Huggins theory.
Here $\epsilon$ and $\sigma$ are, respectively, parameters
fixing the energy and length scales.
Accordingly, our results are reported in
terms of these natural units, using $\tau = \sigma(m/\epsilon)^{1/2}$
for the time scale.  All the results reported here
are obtained using $\Gamma = 0.5\tau^{-1}$ at a constant
overall number density $\rho = 0.85 \sigma^{-3}$ (isochoric) and
a temperature $T = 1.0 \epsilon/k_B$. Chain size $N$ varies from
monomers to 30-mers.

Simply by varying $\delta$, while leaving the temperature constant,
one can induce a phase transition
from a homogeneous blend to a system with
two coexisting phases.
This effect is reminiscent to what is done in
recent experiments reported by Gehlsen {\it et al.}\cite{gehlsen92} in which
the phase separation of binary mixtures
is studied as a function of the difference
in deuterium between two otherwise identical polymer species.
Thus, in principle, the model and an MD simulation method would be adequate
to study the kinetics of phase separation in binary blends.
However, the time scale involved in the
unmixing transition ---occurring
through the diffusive motion of the chains--- is extremely long.
To properly sample the phase space in a reasonable
calculation time, we
supplement the MD moves with an
MC procedure relabeling the type of
the homopolymer chains $(A\leftrightarrow B$)
with a Metropolis transition rule\cite{mc}.
Our exchange attempt rate is approximately $M$ chains per $\tau$\cite{model}.

\begin{figure}[tb]
\centerline{\psfig{figure=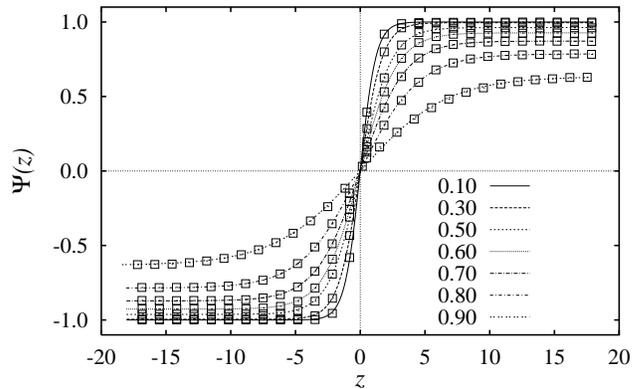,width=\hsize}}
\caption{\label{fig:profiles}
The interfacial profiles $\Psi(z)$ defined
as the $xy$-slice averages of $\psi(\mbox{\bf r})$
for a system of 4000 10-mers with
antiperiodic boundary conditions.
The values of $\delta_c/\delta$ are as indicated.
The lines are from a least-squares fit to the function
$\Psi(z) = \sum_{i=1}^2 a_i \mbox{erf}(\sqrt{\pi}z/w_i)$.
}
\end{figure}

Cubic simulation cells of volume $L^3 = NM/\rho$
and with full periodic boundary conditions
were used to study the bulk properties of our model\cite{model}.
The equilibrium bulk order parameter
$\psi_o(M, N, \delta)$ was measured for different sizes
as a function of the immiscibility parameter $\delta$.
A finite-size scaling analysis\cite{model} of our results, involving
the multiple-histogram technique\cite{deutsch92a}, was used
to extract the critical immiscibility value $\delta_c(N)$
associated with a system of infinite size ($M\rightarrow\infty$).
For $N > 2$, our data is well-described by
$N\delta_c = 3.40(5)k_BT/\epsilon$,
clearly evidencing a $1/N$ behavior of the critical
immiscibility, as predicted by the Flory-Huggins theory.

The finite-size effects on the interfacial
properties of multiphase systems are not as
well-known as for the bulk.
Capillary-wave effects\cite{capillary},
confinement\cite{kerle96}
or finite-chain-size\cite{chaineffects} corrections
to properties such as the interfacial width or
the surface tension are still under active investigation.
We shall now address these questions.

Once the critical miscibility parameter is known,
we can build antiperiodic immiscible
systems (at $\delta_c/\delta < 1$) and
study the interfacial properties of the model,
that is,
we use simulation cells having an antiperiodic\cite{anti,muller}
boundary condition in one direction and a volume $L^2L_\perp$,
where $L_\perp$ is the length perpendicular to the interface.
Figure~\ref{fig:profiles}
shows the interfacial profile of the order parameter $\Psi(z)$
of such systems, obtained by spatially averaging
$\psi(\bbox{r})$ in $xy$ slices,
perpendicular to the antiperiodic direction $z$,
and then averaging the resulting profiles in phase space
with the proper offset (since the zero of the interface is
translationally invariant).
It is tempting to fit the order parameter profile
directly to the same intrinsic interfacial function
$\Psi(z) = \psi_o \tanh(2z/w)$, and thus associate a width
$w$ to the interface. However,
our averaging of the profile incorporates
the effect of the intrinsic interfacial width $w_o$ as well
as any other perturbation broadening
the interface, such as capillary waves.
To separate both effects, a systematic
study of the interfacial profile for different system
sizes has to be performed.

In capillary wave theory\cite{capillary}, the
roughness $\langle (\Delta z_o)^2 \rangle^{1/2}$ of the
zero of a sharp interface
$z_o(x, y)$ is found to be,
in the linearized regime of small distortions,
\begin{equation}
\label{eq:capillary}
\langle (\Delta z_o)^2 \rangle =
\frac{k_BT}{2\pi\gamma}\ln\left(\frac{q_l}{q_L}\right),
\end{equation}
where $q_L$ is the lower cut-off driven by the system size and generally
taken to be $\pi/L$ while $q_l$ is the upper cut-off, usually assumed
to be driven by some correlation length such as $\pi/(c'w_o)$,
where $c'$ is some number.
It is assumed here that $L_\perp \rightarrow \infty$ so that
the roughness is strictly controlled by $L$.

We shall now derive a general and formal argument allowing us to add
the roughening effects of capillary waves to the intrinsic
interfacial width, which we shall measure universally as a moment.
If one assumes that capillary waves can be decoupled from
density fluctuations, the averaged interfacial profile $\Psi(z)$
can be written as the convolution of the intrinsic interfacial
profile $\psi(z-z_o)$, and the probability ${\cal P}(z_o) dz_o$
of finding the interface at $z_o$,
\begin{equation}
\Psi(z) = \int_{-\infty}^{\infty} \psi(z-z_o) {\cal P}(z_o) \; dz_o.
\label{eq:convolution}
\end{equation}
By applying $d/dz$ on each side, one finds that $\Psi'(z)$ is
the convolution of two well-bounded functions, $\psi'(z-z_o)$
and ${\cal P}(z_o)$. We associate a functional
measure $v[f]$ to a well-bounded function $f$ as~\cite{variance}
\begin{equation}
\label{eq:measure}
v[f] \equiv \frac{\int_{-\infty}^{\infty} z^2 f(z) \; dz}%
{\int_{-\infty}^{\infty} f(z) \; dz} =
\frac{-\frac{d^2}{dk^2}\tilde{f}(k)|_{k=0}}%
{\tilde{f}(0)},
\end{equation}
where $\tilde{f}$ is the Fourier transform of $f$.
For real $f$, it can be shown\cite{alex} using the convolution
theorem that $v[\Psi'] = v[\psi'] + v[{\cal P}]$,
which we rewrite as
\begin{equation}
\label{eq:quadrature}
\Delta^2  =
\Delta_o^2 + \frac{k_BT}{2\pi\gamma}\ln\left(\frac{L}{c\Delta_o}\right),
\end{equation}
where $\Delta^2 \equiv v[\Psi']$ now measures the total
interfacial width while $\Delta^2_o \equiv v[\psi']$ is
related to the intrinsic interfacial width $w_o$. The last term
comes from Eq.~\ref{eq:capillary} and the fact that
$v[{\cal P}] = \langle (\Delta z_o)^2 \rangle$.
We reiterate here that
Eq.~\ref{eq:quadrature} is of general validity
since it does not rely on any specific functional form
and only uses well-controlled approximations.
For example, if one fits $\Psi(z)$ to either
a simple $\psi_o \tanh[2z/w]$ or to
$\sum_{i=1}^n a_i \mbox{erf}(\sqrt{\pi}z/w_i)$ \cite{gauss},
then $\Delta^2$ is $\pi^2w^2/48$
or $\sum_{i=1}^n a_iw_i^2/(2\pi\sum_{i=1}^na_i)$,
respectively.

\begin{figure}[tb]
\centerline{\psfig{figure=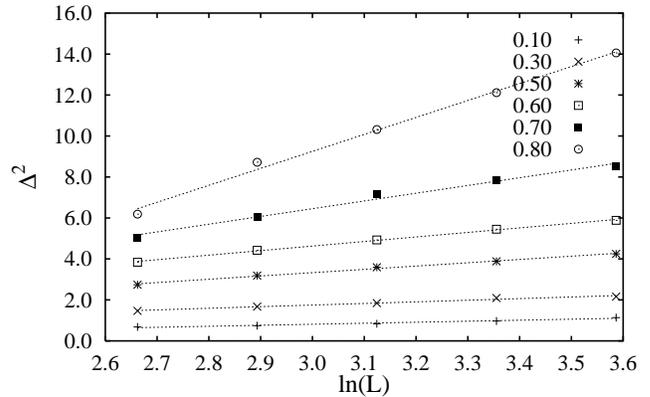,width=\hsize}}
\caption{\label{fig:capillary}
Linear-size and miscibility-parameter dependence
of the interfacial width for various systems of 10-mers.
Values of $\delta_c/\delta$ are as indicated.
The lines are least-squares fits to $\Delta^2 = a_\delta + b_\delta \ln(L)$.
Systems are of size $L^2L_\perp$, with $L_\perp \gg \Delta$.
}
\end{figure}

\begin{figure}[tb]
\centerline{\psfig{figure=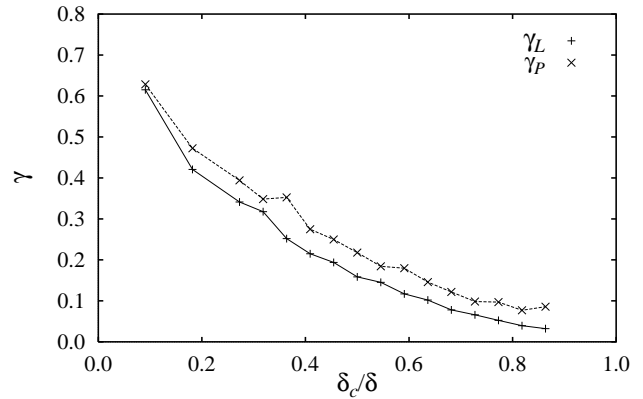,width=\hsize}}
\caption{\label{fig:gammalj}
Two different estimates of the surface tension for
monomeric systems. Surface tension $\gamma_P$ is
obtained from $L_\perp(P_\perp -P_\|)$ while
$\gamma_L = k_BT/(2\pi b_\delta)$,
where $b_\delta$ is obtained from a plot similar
to Fig.~\protect\ref{fig:capillary}.
}
\end{figure}

In Fig.~\ref{fig:capillary}, we present results
from fitting $\Psi(z)$ to $n$=2 error functions.
The dependence of $\Delta^2$ on both the linear
size $L$ of the system (while $L_\perp \gg \Delta$)
and $\delta$ clearly evidences the presence of capillary waves.
To our knowledge,
this is the first time that capillary waves
have been observed in computer simulations
of polymer interfaces\cite{muller}.
We fit our data to the expression
$\Delta^2 = a_\delta + b_\delta\ln(L)$,
where $b_\delta \sim 1/\gamma_L$
(subscript $L$ indicates that the surface tension
has been derived from capillary-wave effects).
By measuring the surface tension $\gamma_P$
independently from the difference
in the pressures (obtained from the virial\cite{Allen_87}, hence
the subscript $P$)
perpendicular and parallel to the interface\cite{Hill_86},
$
\gamma_P = (P_\perp - P_\|)L_\perp,
$ 
one gets another estimate of the surface tension,
and can compare it unambiguously with $\gamma_L$\cite{gammas}.
Figure~\ref{fig:gammalj} shows the two
different measures of $\gamma$ for
a monomeric system. For all $N$, our data show that
with $\gamma_L = k_BT/(2\pi b_\delta)$, both measures are consistent,
thus evidencing the origin of the broadening as being capillary waves.

\begin{figure}[tb]
\centerline{\psfig{figure=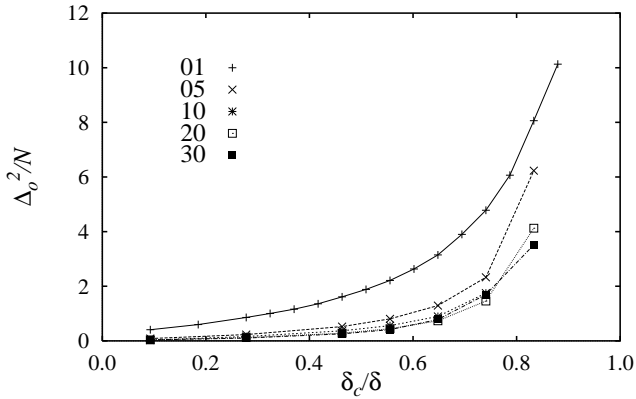,width=\hsize}}
\caption{\label{fig:wo}
The scaled intrinsic interfacial width $\Delta_o^2/N$ as estimated from
Eq.~(\protect\ref{eq:quadrature}) and $c = 13$.. The chain length varies
from monomers to 30-mers.
The data are plotted
so that the reduced miscibility $\delta_c/\delta$ is
equivalent for all chain lengths.
}
\end{figure}

The other parameters $a_\delta$ obtained from the
least-squares fits are always negative. 
Once we know $a_\delta$ and $b_\delta$,
Eq.~\ref{eq:quadrature} can be solved
numerically for $\Delta_o$. It turns out that
our data set only has a solution for $c\gtrsim 13$,
thus imposing a bound on cut-off $q_l$.
Figure~\ref{fig:wo} shows values of $\Delta_o$ extracted by assuming
$c=13$ and plotted such as to compare with
theory which predicts that $w_o^2/N \sim 1/(N\chi_{AB})$,
where $N\chi_{AB}$ is the measure of immiscibility
in the Flory-Huggins theory
(with the critical value $(N\chi_{AB})_c = 2$),
such that $2/(N\chi_{AB})$ corresponds to $\delta_c/\delta$ for our model.

\begin{figure}[tb]
\centerline{\psfig{figure=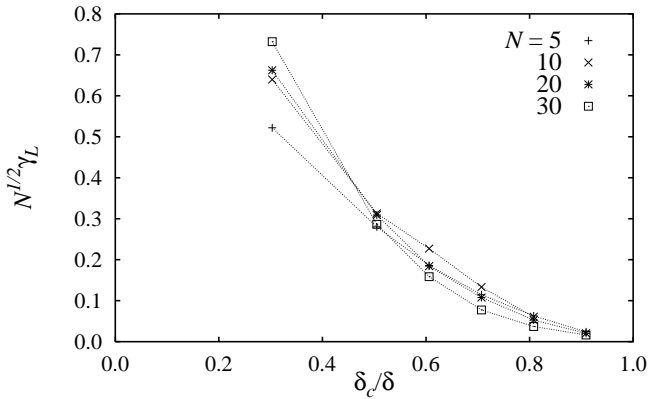,width=\hsize}}
\caption{\label{fig:gammaraw}
The miscibility-parameter dependence
of the surface tension for systems of different chain lengths.
The scaling form holds very well over the range studied.
}
\end{figure}

The $L_\perp$-dependence of the interfacial width has also
been investigated. For $L_\perp \gtrsim 4\Delta$
(and $\delta_c/\delta \lesssim 0.8$),
we find no effect of $L_\perp$ on $\Delta$,
in contrast with recent studies\cite{kerle96}
that observe that $\Delta \sim L_\perp^{1/2}$
for near-critical immiscibilities.
In these studies however,
$L_\perp$ is such that the interface is confined
so that the tails of the order parameter
do not relax to the bulk value.
In the present work, this effect is taken care of
by only using results from systems for
which $L_\perp$ is large enough so that
there exists a wide region
(going through the antiperiodic boundary)
in which $\psi$ is
fully relaxed at $\pm \psi_o(\delta)$.

In Fig.~\ref{fig:gammaraw}, the effect of the
chain length on the surface tension
is shown.
From self-consistent field results
(and as $N\rightarrow \infty$),
$N^{1/2}\gamma \sim (N\chi_{AB})^{1/2}$.
Accordingly, our result collapse very well when
$N^{1/2}\gamma$ is plotted in terms of $\delta_c/\delta$.

We have shown that a continuous-space model with a
rather simple potential allows us to study the
interface between immiscible binary blends of homopolymers.
The natural isotropy of an off-lattice model
coupled with a weak mismatch between the polymer chains
made it possible for us to observe strong capillary-wave effects
at the interface of moderately small systems and evaluate
the surface tension unambiguously.

We would like to thank Scott Milner
for stimulating discussions and
le {\it Fonds FCAR du Qu\'ebec\/} for financial support.


\begin{thebibliography}{99}
\vspace*{\refspace}

\bibitem{meanfield}
J.W. Cahn and J.E. Hilliard,  J. Chem. Phys. {\bf 28}, 258 (1958).

\bibitem{helfand}
E. Helfand and Y. Tagami, J. Chem. Phys. {\bf 56}, 3592 (1972);
M. Lifschitz, K.F. Freed, and H. Tang, \jcp {\bf 103}, 3767 (1995).

\bibitem{Flory_53}
P. Flory, {\em Principles of Polymer Chemistry}
(Cornell University Press, Ithaca, 1953). 

\bibitem{definition}
This definition of $w_o$ is such
that $(\frac{d\psi}{dz})_{z=z_o} \equiv 2\psi_o/w_o$.

\bibitem{experiments}
M.L. Fernandez {\it et al.}, Polymer {\bf 29}, 1923 (1988);
K.R. Shull, A.M. Mayes, and T.P. Russel, Macromolecules {\bf 27},
2732 (1993).

\bibitem{sferrazza97}
M. Sferrazza {\it et al.}, \prl {\bf 78}, 3693 (1997).

\bibitem{capillary}
F.P. Buff, R.A. Lovett, and F.H. Stillinger, \prl {\bf 15}, 621 (1965);
D. Beysens and  M. Robert, \jcp {\bf 87}, 3056, (1987);
A.N. Semenov, {\em Macromolecules} {\bf 27}, 2732 (1994).

\bibitem{chaineffects}
D.W. Schubert and M. Stamm, Europhys. Lett. {\bf 35}, 419 (1996);
D. Broseta {\it et al.}, Macromolecules {\bf 23}, 132 (1990);
M. Lifschitz and K.F. Freed, \jcp {\bf 98}, 8994 (1993);
E. Helfand, S.M. Bhattacharjee, and G.H. Fredrickson,
\jcp {\bf 91}, 7200 (1989).

\bibitem{model}
G.S. Grest {\it et al.}, \jcp {\bf 104}, 10583 (1996);
M.-D. Lacasse and G.S. Grest,
in {\it Morphological Control in Multiphase Polymer Mixtures\/},
edited by R.~M. Briber, D.~G. Pfeiffer, and C.~C. Han
(Material Research Society, Pittsburgh, 1997) Vol. 461, p.\ 129.

\bibitem{Binder_95}
For recent reviews see
{\em Monte-Carlo and Molecular-Dynamics
Simulations in Polymer Science}, edited by K. Binder
(Oxford University Press, New York, 1995).

\bibitem{Hill_86}
T.L. Hill, {\em An Introduction to Stastistical Thermodynamics}
(Dover, New York, 1986) p.\ 316.

\bibitem{robert83}
M. Robert, \prl {\bf 34}, 444 (1983).

\bibitem{Allen_87}
M.P. Allen and D.J. Tildesley, {\em Computer Simulation of Liquids}
(Clarendon, Oxford, 1987).

\bibitem{gehlsen92}
M.D. Gehlsen {\it et~al.}, Phys. Rev. Lett. {\bf 68},  2452  (1992).

\bibitem{mc}
This procedure was first introduced in lattice simulations
and has been very successful. See
A. Sariban and K. Binder, J. Chem. Phys. {\bf 86},  5859  (1987);
Macromolecules {\bf 21}, 711 (1988).

\bibitem{deutsch92a}
H.-P. Deutsch, J. Stat. Phys. {\bf 67},  1039  (1992).

\bibitem{kerle96}
T. Kerle, J. Klein, and K. Binder, \prl {\bf 77}, 1318 (1996).

\bibitem{anti}
An antiperiodic boundary condition (ABC) acts like a
periodic one except that the type is inversed ($A \leftrightarrow B$)
when mers move or interact through the ABC,
thus forcing the existence of at least one interface.

\bibitem{muller}
M. M\"uller, K. Binder, and W. Oed,
J. Chem. Soc. Faraday Trans. {\bf 91}, 2369 (1995);
A. Werner {\it et.\ al}, unpublished (1997).

\bibitem{variance}
If $f$ is normalized then $v[f]$ is simply the variance of $f$.

\bibitem{alex}
A.J. Levine and M.-D. Lacasse, to be published.

\bibitem{gauss}
This choice of functions is equivalent to fitting $\Psi'(z)$ to
either sech$^2(2z/w)$ or $n$ Gaussians.

\bibitem{gammas}
More correctly however, $\gamma_L$ is the energy per interfacial
area while $\gamma_P$ is the energy per cross-sectional area.

\end{thebibliography}
\end{document}